\documentclass[preprint,amsmath,amssymb,aps,pra]{revtex4-2}

\usepackage[utf8]{inputenc}
\usepackage{graphicx}
\usepackage{float}
\usepackage{dcolumn}
\usepackage{caption}
\usepackage{bm}
\usepackage{amsmath}

\usepackage{color}
\usepackage{upgreek}

\newcolumntype{L}[1]{>{\raggedright\let\newline\\\arraybackslash\hspace{0pt}}m{#1}}
\newcolumntype{C}[1]{>{\centering\let\newline\\\arraybackslash\hspace{0pt}}m{#1}}
\newcolumntype{R}[1]{>{\raggedleft\let\newline\\\arraybackslash\hspace{0pt}}m{#1}}

\newcommand{\cm}{cm$^{-1}$ }

\newcommand{\threej}[6]{
\begingroup
\footnotesize{
\renewcommand*{\arraystretch}{0.6}
\begin{pmatrix}
   #1 & #2 & #3 \\
   #4 & #5 & #6 
\end{pmatrix}}
\endgroup
}

\begin{document}

\title{An efficient and flexible approach for computing rovibrational polaritons from first principles}

\author{Tam\'as Szidarovszky}
    \email{tamas.janos.szidarovszky@ttk.elte.hu}
    \affiliation{%
    Institute of Chemistry,
	ELTE E\"otv\"os Lor\'and University, H-1117 Budapest, P\'azm\'any P\'eter s\'et\'any 1/A, Hungary}%


\begin{abstract}
A theoretical framework is presented for the computation of rovibrational polaritonic states of a molecule in a lossless infrared (IR) microcavity.
In the proposed approach the quantum treatment of the rotational and vibrational motion of the molecule can be formulated using arbitrary approximations.
The cavity-induced changes in electronic structure are treated perturbatively, which allows using the existing polished tools of standard quantum chemistry for determining electronic molecular properties.
As a case study, the rovibrational polaritons and related thermodynamic properties of H$_2$O in an IR microcavity are computed for varying cavity parameters and applying various approximations to describe the molecular degrees of freedom.
The self-dipole interaction is found to be significant for nearly all light-matter coupling strengths investigated, and the molecular polarizability proved to be important for the correct qualitative behavior of the energy level shifts induced by the cavity. On the other hand, the magnitude of polarization remains small, justifying the perturbative approach for the cavity-induced changes in electronic structure.
Comparing results obtained using a high-accuracy variational molecular model with those obtained utilizing the rigid rotor and harmonic oscillator approximations revealed that
as long as the rovibrational model is appropriate for describing the field-free molecule, the computed rovibropolaritonic properties can be expected to be accurate as well.
Strong light-matter coupling between the radiation mode of an IR cavity and the rovibrational states of H$_2$O lead to minor changes in the thermodynamic properties of the system, and these changes seem to be dominated by non-resonant interactions between the quantum light and matter.

\end{abstract}

\maketitle

\section{\label{introduction}Introduction}

As an alternative to interactions of molecules with intense laser fields, strong light-matter coupling can also be achieved by the confinement of molecules in microscale optical or plasmonic cavities.
When the light-matter coupling is stronger than the loss of the cavity mode and the decay rates in the matter, the strong coupling regime is reached, leading to the formation of hybrid light-matter states, called polaritons
\cite{Ebbesen_review_2016,Zhou_review_2018,Feist_GarciaVidal_ACSphotonics_2018,Edina1,Herrera_PRL_2016,Hertzog2019,Herrera_2020,Kowalewski2017}.
The confined photonic modes of the cavity can efficiently couple with either electronic or vibrational molecular states, depending on the cavity mode wavelength.
The formation of vibrational polaritons in infrared (IR) microcavities has received a great deal of attention \cite{Hirai_2020_review,Nagarajan2021,Simpkins2021,Wang2021,Sidler2022,Li2022,Dunkelberger2022}, owing to the possibility of modifying thermal chemical reactions with vibrational polaritons.
Despite promising progress \cite{Schfer2022,Schfer2022b},
the full understanding of the effects of vibrational strong coupling and the formation of vibrational polaritons on chemical reactivity is a challenge yet to be tackled, as detailed in the reviews cited above.

In addition, closely related to chemical reactivity, the spectroscopy of vibrational polaritons have also been studied both by means of experiments \cite{Shalabney2015,Fleischer_2019,Xiang_2020,Yang2020,Xiang2019,Xiang2018,Dunkelberger_2016,Grafton_2021, Takele_2021,Wright2023} and theory \cite{Strashko2016,Saurabh2016,FRibeiro2018,Hernandez_2019,Triana_2020,Ribeiro_2021,Fischer_2021,Yu2022,Bonini2022,Szidarovszky2021}.
For examples, a detailed investigation of the vibrational polaritons of single anharmonic vibrational modes was carried out in Refs. \cite{Hernandez_2019,Triana_2020}, while Ref. \cite{Ribeiro_2021} showed the presence of unique optical nonlinearities of $N$ anharmonic oscillators interacting with an IR cavity mode.
Using a one-dimensional description Ref. \cite{Fischer_2021} simulated various properties 
of anharmonic oscillators (OH, LiH, NH$_3$), with coordinate dependent dipole functions, coupled to an IR photonic mode, from the weak to ultrastrong coupling regimes.
Refs. \cite{Saurabh2016,FRibeiro2018} provide empirically adjustable anharmonic approaches, which allow incorporating decay mechanisms.
The method of vibrational configuration interaction (VCI) has also been implemented in the context of vibrational polaritons \cite{Yu2022}, and the vibropolaritonic energies and IR spectra of H$_2$O molecules were computed.

Despite the cavity radiation being resonant with (ro)vibrational transitions, the radiation-induced changes in the electronic structure might also become important \cite{Bonini2022}.
For IR cavities, in which the photon energy and respective frequency is comparable to those in rovibrational transitions, the molecule can be assumed to have an electronic structure adiabatically transformed by the radiation field, known as the cavity Born-Oppenheimer approximation (CBO) \cite{Flick2017_pnas,Flick2017_jctc}.
To describe nuclear dynamics, CBO combines naturally with the QEDFT \cite{Tokatly2013,Ruggenthaler2014,Malave2022,Schfer2021} and QEDCC \cite{Haugland2020,Haugland2021} methods, whereby the effect of the cavity radiation is explicitly included in the electronic structure calculation, and the nuclear and photonic modes are governed by the resulting CBO potential energy surfaces.
Using QEDFT and linear response theory, the vibrational polaritons of CO$_2$ and Fe(CO)$_5$ molecules could be computed from first principles \cite{Bonini2022}.

In this work we present an alternative theoretical framework to compute the (ro)vibrational polaritons of molecules in lossless IR microcavities. The proposed approach
is applicable as long as the molecule can be accurately described with a single electronic state, i.e., the Born-Oppenheimer (BO) approximation holds, the radiation field is not resonant with electronic transitions and the ultrastrong coupling regime is not reached.
The presented framework
extends existing computational tools in three main ways:
(1) the exact quantum treatment of molecular rotation is incorporated, which can be important considering that the coupling strength between the transition dipole and the radiation field, having laboratory-fixed polarization, depends on molecular orientation; (2)
the cavity-induced changes in electronic structure are treated perturbatively, which allows using the existing efficient tools of standard quantum chemistry and theoretical molecular spectroscopy to determine electronic molecular properties; and (3) various approximations and levels of theory regarding the description of the molecular degrees of freedom and light-matter interaction can be made.
The flexibility of the proposed framework can be useful to arrange the appropriate balance between accuracy and computational effort, allowing it to be used for a wide range of molecules of various complexity.

\section{\label{Theory}Theoretical foundations}

\subsection{Rovibrational polaritons}

The Hamiltonian of a molecule interacting with a
single lossless cavity mode can be written in the dipole approximation
as a sum of the field-free molecular rovibronic Hamiltonian $(\hat{H}_{\rm m}^{\rm tot})$, the radiation mode Hamiltonian $(\hat{H}_{{\rm c}})$, the interaction term between the molecular dipole and the electric field of the radiation mode $(-\hat{{\rm \mathbf{E}}}_{{\rm c}}\hat{\mathbf{\upmu}}^{\rm tot})$, and the self-dipole energy term $(\hat{H}_{\rm sd})$ \cite{CohenTannoudji,Rokaj2018,Schfer2020}
 \begin{equation}
 \begin{aligned}
\hat{H}&=\hat{H}_{\rm m}^{\rm tot} +\hat{H}_{{\rm c}} -\hat{{\rm \mathbf{E}}}_{{\rm c}}\hat{\mathbf{\upmu}}^{\rm tot} + \hat{H}_{\rm sd} \\
&= \hat{H}_{{\rm m}}^{\rm tot} + \hbar\omega_{\rm c}\hat{a}_{{\rm c}}^{\dagger}\hat{a}_{{\rm c}} -\sqrt{\frac{\hbar\omega_{{\rm c}}}{2\varepsilon_{0}V}}\mathbf{e}\hat{\mathbf{\upmu}}^{\rm tot}(\hat{a}_{{\rm c}}^{\dagger}+\hat{a}_{{\rm c}}) + \frac{1}{2\varepsilon_{0}V} (\mathbf{e}\hat{\mathbf{\upmu}}^{\rm tot})^2 \\
&= \hat{H}_{{\rm m}}^{\rm tot} + \hbar\omega_{\rm c}\hat{a}_{\rm c}^{\dagger}\hat{a}_{\rm c} - \frac{g}{ea_{0}} \mathbf{e}\hat{\mathbf{\upmu}}^{\rm tot}(\hat{a}_{{\rm c}}^{\dagger} + \hat{a}_{{\rm c}}) +\Big(\frac{g}{ea_{0}}\Big)^2 \frac{1}{\hbar\omega_{{\rm c}}}(\mathbf{e}\hat{\mathbf{\upmu}}^{\rm tot})^2
\label{eq:hamiltonian_full_1}
 \end{aligned}
 \end{equation}
where $\hat{a}_{c}^{\dagger}$ and $\hat{a}_{c}$
are photon creation and annihilation operators, respectively, $\omega_{c}$
is the frequency of the cavity mode, $\hslash$ is Planck's constant
divided by $2\pi$, $\varepsilon_{0}$ is the electric constant, $V$
is the volume of the electromagnetic mode, $\mathbf{e}$ is the polarization
vector of the cavity mode, $\hat{\mathbf{\upmu}}^{\rm tot}$ is the total
dipole moment operator of the molecule,
$e$ is the elementary charge, and $a_{0}$ is the Bohr radius. In the last line of Eq. (\ref{eq:hamiltonian_full_1}) the quantity
$g=e a_{0} \sqrt{\hbar\omega_{{\rm c}}/(2 \varepsilon_{0}V)}$ was introduced, which represents the coupling strength between a single photon electric field and the atomic unit of the dipole moment.

As already mentioned in the introduction, for IR cavities, in which the photon energy is comparable to those in rovibrational transitions, the cavity-induced changes in the electronic structure can be incorporated into the CBO potential energy surfaces, which govern the nuclear and photonic degrees of freedom.
In this work the effects of the cavity radiation on the electronic structure is accounted for perturbatively, using the electronic polarizability.
Such a perturbative approach was also applied in Ref. \cite{Galego_2019}, and is a common practice for electronically non-resonant laser fields of moderate intensity \cite{03StSe,19KoLeSu,Szidarovszky2017,18SzJoYa,Simk2022}.
Therefore, for IR cavity radiations, being off resonant with the electronic transitions, and for coupling strengths tipically realized in experiments \cite{Xiang2018,Wright2023}, the perturbative treatment of electronic excitation seems to be a reasonable approach.
Note that additional care should be taken if the simulations involve nuclear configurations where the energy separation of potential energy surfaces (PES) are small. Such situations can lead to significant nonadiabatic couplings \cite{Worth2004} or electronic resonance with the IR radiation, giving rise to light-induced conical intersections \cite{Fbri2021,Szidarovszky2018}, both of which require approaches beyond the one presented here \cite{Worth2004,Schfer2018,Fbri2020}.

Taking the expectation value of the Hamiltonian in Eq.(\ref{eq:hamiltonian_full_1}) with the cavity-modified electronic wave function, and utilizing the standard approach \cite{06BuJe} of expressing the modified dipole with the field-free permanent dipole $\hat{\mathbf{\upmu}}_0$ and the first-order static polarizability $\hat{\mathbf{\alpha}}$, both nuclear coordinate-dependent, one obtains the following rovibrophotonic Hamiltonian.
\begin{equation}
\begin{aligned}
\hat{H}&=
\hat{H}_{{\rm rovib}}
+\hbar\omega_{{\rm c}}\hat{a}_{{\rm c}}^{\dagger}\hat{a}_{{\rm c}}
 -\frac{g}{ea_{0}} \mathbf{e}\hat{\mathbf{\upmu}}_0(\hat{a}_{{\rm c}}^{\dagger}+\hat{a}_{{\rm c}})\\
&-\frac{(g/ea_{0})^2}{2}\mathbf{e}\hat{\mathbf{\alpha}}\mathbf{e}(\hat{a}_{{\rm c}}^{\dagger}+\hat{a}_{{\rm c}})(\hat{a}_{{\rm c}}^{\dagger}+\hat{a}_{{\rm c}})
 + \frac{(g/ea_{0})^2}{\hbar\omega_{{\rm c}}}(\mathbf{e}\hat{\mathbf{\upmu}}_0)^2\\
&+\frac{(g/ea_{0})^3}{2\hbar\omega_{{\rm c}}}\big[ (\mathbf{e}\hat{\mathbf{\upmu}}_0)(\mathbf{e}\hat{\mathbf{\alpha}}\mathbf{e})+(\mathbf{e}\hat{\mathbf{\alpha}}\mathbf{e})(\mathbf{e}\hat{\mathbf{\upmu}}_0) \big](\hat{a}_{{\rm c}}^{\dagger}+\hat{a}_{{\rm c}})+\frac{(g/ea_{0})^4}{4\hbar\omega_{{\rm c}}}(\mathbf{e}\hat{\mathbf{\alpha}}\mathbf{e})^2(\hat{a}_{{\rm c}}^{\dagger}+\hat{a}_{{\rm c}})^2,
\end{aligned}
\label{eq:hamiltonian_full_2}
\end{equation}
where $\hat{H}_{{\rm rovib}}$ is the field-free rovibrational Hamiltonian utilizing the unperturbed electronic PES. For the sake of simplicity, from now on we omit the terms with $g^3$ and $g^4$ (for the H$_2$O molecule and cavity parameters investigated in this work, it was numerically tested that these terms are in fact negligible).
The cavity mode is assumed to be the one with the lowest photon energy, including additional overtones is expected not to affect the numerical results presented below. Extending the presented approach with further cavity modes is in principle straightforward, and could become important, for example if different modes are resonant with different (ro)vibrational excitations, but this is beyond the scope of this proof-of-concept work.
To describe the rovibropolaritonic system, the $\vert N\rangle \vert \Psi^{nJM}\rangle$ direct-product functions provide a complete
basis, where the $\vert\Psi^{nJM}\rangle$ field-free
rovibrational eigenstates satisfy 
\begin{equation}
\hat{H}_{{\rm rovib}}\vert\Psi^{nJM}\rangle=E^{nJ}\vert\Psi^{nJM}\rangle,\label{eq:field_free_schr_Eq}
\end{equation}
$J$ and $M$ are the rotational angular momentum and its projection
onto the space-fixed z-axis, respectively, $n$ is all other quantum
numbers uniquely defining the rovibrational states, and $\vert N\rangle$
is a photon number state of the cavity radiation.
The matrix representation of Eq. (\ref{eq:hamiltonian_full_2}) using the field-free rovibrational eigenstates $\vert\Psi^{nJM}\rangle$ as molecular basis functions can be expressed conveniently by using the spherical basis representation of the molecular dipole and polarizability, which are obtained from the Cartesian representation as \cite{88Zare}
    $\mu^{(1,0)}=\mu_3$,
    $\mu^{(1,\pm1)}=\frac{1}{\sqrt{2}}(\mp \mu_1-i\mu_2)$,
    $\alpha^{(0)}=-\frac{1}{\sqrt{3}}(\alpha_{11}+\alpha_{22}+\alpha_{33})$,
    $\alpha^{(2,\pm 2)}=\frac{1}{2}[\alpha_{11}-\alpha_{22}\pm i (\alpha_{12}+\alpha_{21})]$,
    $\alpha^{(2,\pm 1)}=\frac{1}{2}[\mp(\alpha_{13}+\alpha_{31})-i(\alpha_{23}+\alpha_{32})]$, and
    $\alpha^{(2,0)}=\frac{1}{\sqrt{6}}(2\alpha_{33}-\alpha_{22}-\alpha_{11})$.
Assuming the cavity electric field to be polarized along the lab-fixed z-axis, $\mathbf{e}=(0,0,1)$, one obtains matrix elements of the following form
    \begin{equation}
    \begin{aligned}
   & \langle N \vert \langle \Psi ^{JMn} \vert \hat{H} \vert \Psi ^{J'M'n'} \rangle \vert N' \rangle =
        E ^{Jn} \delta_{JJ'} \delta_{nn'} \delta_{MM'} \delta_{NN'} &\\
    &- \frac{g}{ea_{0}} \Big( \sqrt{N'+1}\delta_{NN'+1}+\sqrt{N'}\delta_{NN'-1}\Big) \sum_{k=-1}^{1} \langle \Psi ^{JMn} \vert {D_{0k}^{1}}^*  \mu ^{{\rm BF},(1,k)} \vert \Psi ^{J'M'n'} \rangle  \\
    &-\frac{(g/ea_{0})^2}{\sqrt{6}} \Big( \sqrt{(N'+1)(N'+2)}\delta_{N,N'+2}+(2N'+1)\delta_{N,N'}+\sqrt{N'(N'-1)}\delta_{N,N'-2}\Big)\times \\
    &\Bigg[ \sum_{k=-2}^{2} \langle \Psi ^{JMn} \vert {D_{0k}^{2}}^*  \alpha ^{{\rm BF},(2,k)} \vert \Psi ^{J'M'n'} \rangle - \frac{1}{\sqrt{2}}\langle \Psi ^{JMn} \vert \alpha ^{{\rm BF},(0)} \vert \Psi ^{J'M'n'} \rangle \Bigg] \\
    &+  \delta_{NN'}\delta_{MM'}\frac{(g/ea_{0})^2}{\hbar\omega_{{\rm c}}} \sum_{k,k'=-1}^{1} \sum_{J'',n''} \langle \Psi ^{JMn} \vert {D_{0k}^{1}}^* \mu ^{{\rm BF},(1,k)} \vert \Psi ^{J''Mn''} \rangle \langle \Psi ^{J''Mn''} \vert   {D_{0k'}^{1}}^* \mu ^{{\rm BF},(1,k')} \vert \Psi ^{J'M'n'} \rangle,
    \end{aligned}
    \label{eq:hamiltonian_in_general_basis}
    \end{equation}
where the relation $\sum_{J'',M'',n''} \vert \Psi ^{J''M''n''} \rangle \langle \Psi ^{J''M''n''} \vert =\hat{I}$ for the molecular state space was used. 
In the above equation the superscript BF stands for body-fixed components, i.e., those directly obtained from quantum chemistry computations, and $D_{km}^{j}$ are the Wigner-D matrices \cite{88Zare} responsible for transforming the body-fixed spherical basis components of the molecular properties into their lab-fixed components (which are used to express the interaction with the cavity electric field in Eq.(\ref{eq:hamiltonian_full_2})).
Using the general variational expansion $\vert \Psi ^{JMn} \rangle = \sum_{K,v}{C^{Jn}_{Kv} \vert v \rangle \vert J K M \rangle}$ for the field-free rovibrational eigenstates, where $\vert v \rangle$ are vibrational basis functions and $\vert J K M \rangle$ are the symmetric top rotational eigenstates \cite{88Zare}, the terms containing the Wigner-D matrices in Eq. (\ref{eq:hamiltonian_in_general_basis}) can be evaluated as
    \begin{equation}
    \begin{aligned}
     \sum_{k=-1}^{1} \langle \Psi ^{JMn} \vert {D_{0k}^{1}}^*  \mu ^{{\rm BF},(1,k)} \vert \Psi ^{J'M'n'} \rangle = \\
   \sum_{k=-1}^{1} \Big( \sum_{v,v'} \langle v \vert \mu ^{{\rm BF},(1,k)} \vert v' \rangle \sum_{K,K'} {C_{Kv}^{Jn}}^* C_{K'v'}^{J'n'} \langle JKM \vert {D_{0k}^{1}}^* \vert J'K'M' \rangle \Big),
    \end{aligned}
    \label{eq:hamiltonian_in_variational_basis_mu0}
    \end{equation}
    and
    \begin{equation}
    \begin{split}
    \sum_{k=-2}^{2} \langle \Psi ^{JMn} \vert {D_{0k}^{2}}^*  \alpha ^{{\rm BF},(2,k)} \vert \Psi ^{J'M'n'} \rangle - \frac{1}{\sqrt{2}}\langle \Psi ^{JMn} \vert \alpha ^{{\rm BF},(0)} \vert \Psi ^{J'M'n'} \rangle = \\
    \sum_{k=-2}^{2} \Big( \sum_{v,v'} \langle v \vert \alpha ^{{\rm BF},(2,k)} \vert v' \rangle \sum_{K,K'} {C_{Kv}^{Jn}}^* C_{K'v'}^{J'n'} \langle JKM \vert {D_{0k}^{2}}^* \vert J'K'M' \rangle \Big) \\ - \delta_{JJ'} \delta_{MM'} \frac{1}{\sqrt{2}} \sum_{v,v'} \langle v \vert \alpha ^{{\rm BF},(0)} \vert v' \rangle \sum_{K}{C_{Kv}^{Jn}}^* C_{Kv'}^{Jn'},
    \end{split}
    \label{eq:hamiltonian_in_variational_basis_alpha}
    \end{equation}
    where $\langle JKM \vert {D_{0k}^{j}}^* \vert J'K'M' \rangle=(2J+1)^{1/2}(2J'+1)^{1/2}(-1)^{-k+M'-K'}\threej{J}{j}{J'}{M}{0}{-M'}\threej{J}{j}{J'}{K}{-k}{-K'}$, which implies that $M=M'$ for all non-zero matrix elements, as expected for a system with cylindrical symmetry.
    The $k$th rovibrational polaritonic $\vert\Psi_{{\rm pol}}^{k,M}\rangle$ eigenstate and $E_{\rm pol}^{k,M}$ energy are computed as the $k$th eigenvector and eigenvalue, respectively, of the Hamiltonian defined by Eq. (\ref{eq:hamiltonian_in_general_basis}). 
    
    The rovibrational polaritons of an H$_2$O molecule in an IR microcavity are simulated using two different molecular models and various approximations within each model.
    In the first molecular model, results of a high-accuracy variational rovibrational simulation are used for constructing the matrix elements in Eq. (\ref{eq:hamiltonian_in_general_basis}).
    This approach can be regarded as the theoretical benchmark within the framework utilized.
    The second molecular model is based on the harmonic oscillator and rigid-rotor (HORR) approximations \cite{06BuJe}, with all parameters taken from black-box quantum chemistry calculations.
    The second model aims to reveal the robustness and flexibility of the approach presented in this work, important for considering applications to larger systems.

\clearpage

\section{Computational details}
All results presented in this work are converged with respect to the size of the basis employed to construct the Hamiltonian in Eq. (\ref{eq:hamiltonian_in_general_basis}). The largest basis used includes a maximum photon number of two, and all rovibrational states with $J\leq10$ and an energy no more than 5500 cm$^{-1}$ above the zero point vibrational enegry (ZPVE). Such parameters ensure that all vibrational fundamentals and the bending overtone of H$_2$O, including rotationally excited states, are included in the basis.
The two methods for obtaining the field-free molecular rovibrational eigenstates, to be used in the direct product basis employed in Eq. (\ref{eq:hamiltonian_in_general_basis}), are detailed below.

\subsection{Molecular model I.} In the high-accuracy variational approach, the field-free, bound rovibrational eigenstates of H$_2$O were computed using the D2FOPI protocol and program suite \cite{10SzCsCz}, with the PES of Ref. \cite{18PoKyZoTe}.
In these computations the C$_{\rm 2v}$(M) molecular symmetry \cite{06BuJe} was kept as described in Ref. \cite{13SzCs}.
In the D2FOPI program, the time-independent rovibrational Schrödinger equation for a triatomic molecule is solved by an iterative eigensolver using symmetry adopted Wigner matrices \cite{88Zare} as rotational basis functions and a mixed discrete variable representation (DVR) \cite{00LiCa} and finite basis representation \cite{00LiCa} along the vibrational degrees of freedom expressed in the orthogonal Jacobi coordinate system.

For each irreducible representation of the C$_{\rm 2v}$(M) molecular symmetry group, the calculations included (1) a complete set of rotational basis functions, whose size varies depending on the given value of the $J$ rotational quantum number, (2) 45 potential optimized (PO) spherical DVR basis functions \cite{10SzCsCz} along the $R_1\equiv R$(H-H) coordinate, (3) 55 PO spherical DVR basis functions along the $R_2\equiv R$(O-H$_2$) coordinate, connecting the O atom with the center of mass of the H$_2$ moiety, and (4) 25 associated Legendre functions along the $\theta$ angle coordinate defined by the directions of the two stretching coordinates.
The coordinate ranges used in the calculations were R(H-H)/bohr$\in$(0,10) and R(O-H$_2$)/bohr$\in$(0,6), and the nuclear masses $m_{\rm H}$=1.00727647 u and $m_{\rm O}$=15.990526 u were employed.

With these parameters the D2FOPI computations provide the $C_{Kv}^{Jn}$ rovibrational eigenvector coefficients (see Eqs. (\ref{eq:hamiltonian_in_variational_basis_mu0}) and (\ref{eq:hamiltonian_in_variational_basis_alpha})) and the $E^{Jn}$ rovibrational energies (used in Eq. (\ref{eq:hamiltonian_in_general_basis})), all values used in this work converged to within 0.01 cm$^{-1}$.
The obtained vibrational fundamentals are $\tilde{\nu}_1=$3657.05 cm$^{-1}$, $\tilde{\nu}_2=$1594.73 cm$^{-1}$, and $\tilde{\nu}_3=$3796.99 cm$^{-1}$.
The dipole moment surfaces $\mu ^{{\rm BF} ,(1,k)}(R_1,R_2,\theta)$ and the polarizability surfaces $\alpha ^{{\rm BF},(0)}(R_1,R_2,\theta)$ and $\alpha ^{{\rm BF},(2,k) }(R_1,R_2,\theta)$ were generated using the results of Refs. \cite{Lodi2011} and \cite{Avila2005}, respectively.
The matrix elements $\langle v \vert \mu^{{\rm BF},k} \vert v' \rangle$ of Eq. (\ref{eq:hamiltonian_in_variational_basis_mu0}) and $\langle v \vert \alpha ^{{\rm BF},(2,k)} \vert v' \rangle$ and $\langle v \vert \alpha ^{{\rm BF},(0)} \vert v' \rangle$ of Eq. (\ref{eq:hamiltonian_in_variational_basis_alpha}) can be efficiently computed with the DVR basis employed in D2FOPI \cite{Szidarovszky2017}. 

\subsection{Molecular model II.}
The second molecular model employed in this work uses the HORR approximation, with molecular parameters obtained from black-box quantum chemistry computations, carried out with MOLPRO \cite{MOLPRO} on the gold standard CCSD(T)/aug-cc-pVQZ level \cite{89RaTrPoHe,89Dunning}.
The obtained harmonic frequencies and rotational constants are $\tilde{\nu}_1=3830.84$ cm$^{-1}$, $\tilde{\nu}_2=1649.72$ cm$^{-1}$, $\tilde{\nu}_3=3940.43$ cm$^{-1}$, $B_z=14.5720$ cm$^{-1}$, $B_y=9.4934$ cm$^{-1}$ and $B_x=27.2393$ cm$^{-1}$, where the molecule is in the body-fixed xz plane, with the z-axis pointing towards the oxygen. The body-fixed $\mu ^{\rm BF}_{i}(Q_1,Q_2,Q_3) $ dipole and $\alpha ^{\rm BF}_{i,j}(Q_1,Q_2,Q_3)$ polarizability functions were expanded along the normal coordinates to second order, using finite differences to compute their derivatives with respect to the normal coordinates, then transformed to the spherical representation for Eqs. (\ref{eq:hamiltonian_in_variational_basis_mu0}-\ref{eq:hamiltonian_in_variational_basis_alpha}).
Due to this expansion along the normal coordinates, the 
matrix elements $\langle v \vert \mu^{{\rm BF},k} \vert v' \rangle$ of Eq. (\ref{eq:hamiltonian_in_variational_basis_mu0}) and $\langle v \vert \alpha ^{{\rm BF},(2,k)} \vert v' \rangle$ and $\langle v \vert \alpha ^{{\rm BF},(0)} \vert v' \rangle$ of Eq. (\ref{eq:hamiltonian_in_variational_basis_alpha}) can be determined with analytical formulae in the HO vibrational basis.
In the HORR approximation the $C^{Jn}_{Kv}$ expansion coefficients of Eqs. (\ref{eq:hamiltonian_in_variational_basis_mu0}-\ref{eq:hamiltonian_in_variational_basis_alpha}) simplify to $C^{Jn}_{K}$, and can be obtained by solving the RR problem in the $\vert JKM \rangle$ rotational basis. 

\clearpage

\section{Results and discussion}

\subsection{Rovibrational polaritonic energies}
\begin{figure}[h!]
\centering \includegraphics[width=1.00\textwidth]{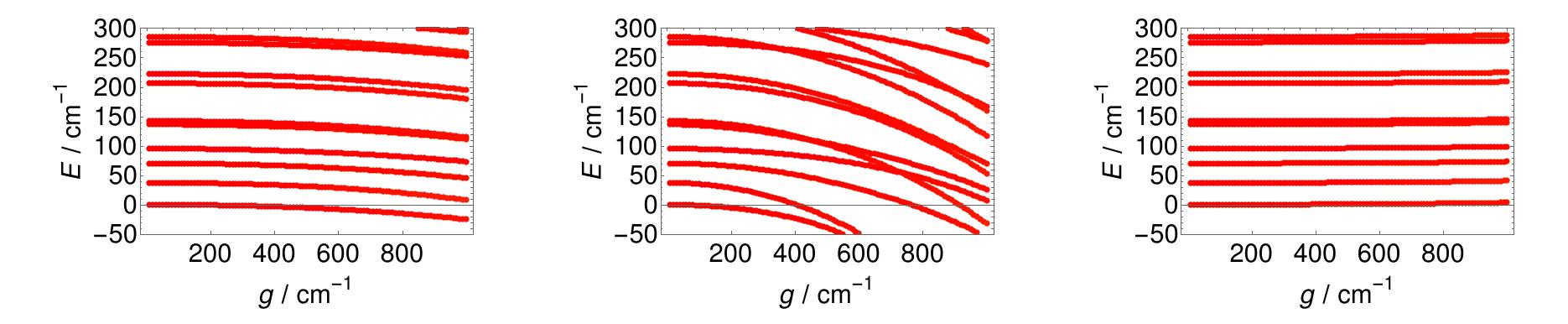}
\centering \includegraphics[width=1.00\textwidth]{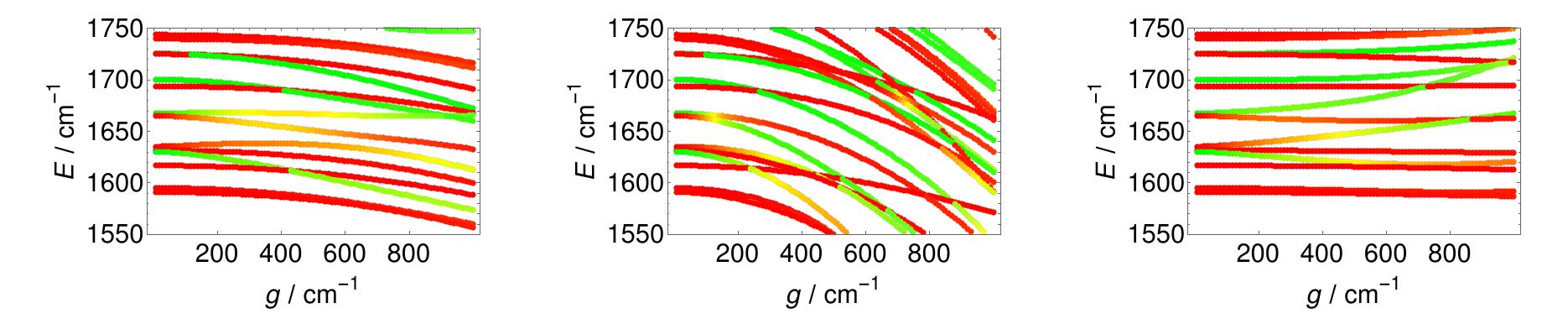}
\centering \includegraphics[width=1.00\textwidth]{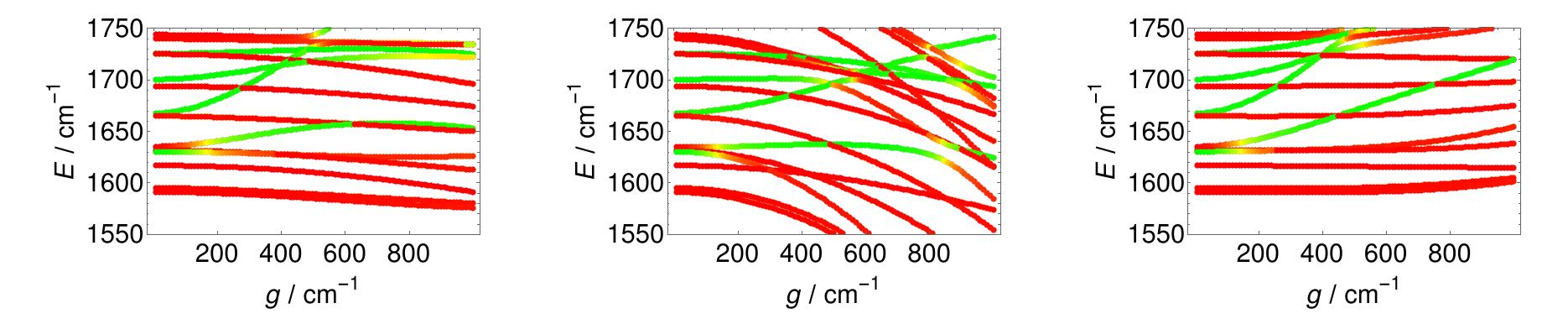}

\caption{Left panels: converged eigenvalues of the rovibrophotonic Hamiltonian of Eq. (\ref{eq:hamiltonian_in_general_basis}) as a function of the $g$ coupling strength, obtained using the numerically exact variational molecular eigenstate basis and a photon wave number of $\tilde{\nu}=1630$ cm$^{-1}$. Middle panels: same as in left panels, but with the self-dipole interaction neglected. Right panels: same as
left panels, but with the induced dipole (polarizability) interaction neglected.
Upper and middle rows were obtained with a maximum photonnumber of two, while the bottom row was obtained with a maximum photon number of one.
The colors of the lines represent their character: red indicates zero expectation value for the photon number, while green represents one photon expectation value. Yellow indicates a mixture of photonic and material excitations.
}
. \label{fig:gvar_var_v2}
\end{figure}

Figure \ref{fig:gvar_var_v2} shows the computed eigenvalues of the Hamiltonian in Eq. (\ref{eq:hamiltonian_in_general_basis}) as a function of the $g$ coupling strength, using the variational molecular eigenstates in the basis and a photon energy of 1630 cm$^{-1}$, nearly resonant with the HOH bending fundamental, i.e., the $(0 1 0)[1 1 1]\leftarrow(0 0 0)[0 0 0]$ rovibrational transition at 1635.0 cm$^{-1}$ (note that the $\Delta J=0$ purely vibrational transition is forbidden).
The middle panels of Figure \ref{fig:gvar_var_v2} were obtained by neglecting the self-dipole term from the Hamiltonian, while the right panels were generated by dropping the polarizablity.
As seen in Figure \ref{fig:gvar_var_v2}, for the smallest $g$ values both the molecular polarizability and the self-dipole interactions are negligible, as expected, although the effects of both start to become significant already around $g=200$ cm$^{-1}$. The magnitude of the self-dipole energy proves to be larger than the energy shifts due to polarizability, although the latter is also important to obtain the qualitatively correct negative energy shifts with increasing coupling strength (compare left and right panels of Figure \ref{fig:gvar_var_v2}). At the largest coupling strength, the energy shifts due to polarizability reach only around a few tens of cm$^{-1}$, which indicates that higher-order polarizabilities are not needed for the coupling strengths investigated here, and that the perturbative inclusion of the cavity-induced changes in the electronic structure of the molecule is appropriate.
Note that in principle the negative energy shifts due to polarizability could be positive due to the self-polarization (see last line of Eq. (\ref{eq:hamiltonian_full_2})) \cite{Schfer2020}.
However, the overall energy shift is not only caused by an isotropic polarization of the electronic wave function, but also by the alignment of the different molecule-fixed axes  with respect to the polarization vector of the cavity electric field. For zero light-matter coupling the wave function shows an isotropic distribution, but with increasing coupling strength the different rotational states become coupled, and the different polarizability along the different molecular axes induces alignment \cite{03StSe}.
This is an important aspect of the employed molecular model, which fully accounts for molecular rotations.

The first row of Figure \ref{fig:gvar_var_v2} shows that on the energy scale of moderate rotational excitation there is no polariton formation, as expected, because the IR cavity photon is far from resonance. Only minor energy shifts are seen at large $g$ coupling strengths.
These observations also indicate that it is safe to neglect further cavity modes (overtones of the $\tilde{\nu}=1630$ cm$^{-1}$ mode) if the investigated energy range is limited to the bending fundamental and below.
The middle and bottom rows of Figure \ref{fig:gvar_var_v2} also demonstrate that molecular energy shifts due to the light-matter interaction have a strong effect on polariton formation, as these level shifts can create or destroy the resonance condition between photon-excited or rovibrationally excited states.
Depending on the relative polarizability of the molecule in the photon-excited state with respect to the rovibrationally excited state,
the efficient generation of polaritons at low coupling strengths could require the photon energy to be either red- or blue-shifted with respect to the field-free molecular transition.
It can also be seen in Figure \ref{fig:gvar_var_v2} that polariton formation (the plotted lines becoming yellow) does not occur for many resonant crossings between the one-photon (green) and zero-photon (red) curves, demonstrating that the transitions between the molecular components of most eigenstates are optically forbidden.
On the other hand, comparing the bottom and middle rows of Figure \ref{fig:gvar_var_v2} shows that the $g$ parameter range of polariton formation is greatly enhanced if two-photon states are also included in the basis.
This can be explained by considering that including two-photon states introduces their one-photon coupling with the one-photon-excited manifold, as well as their two-photon coupling with the zero-photon manifold through polarizability. Such additional couplings lead to the mixing of the uncoupled states, easing optical selection rules through intensity borrowing \cite{Szidarovszky2019}, thus enhancing polariton formation.

\begin{figure}[h!]
\centering \includegraphics[width=1.00\textwidth]{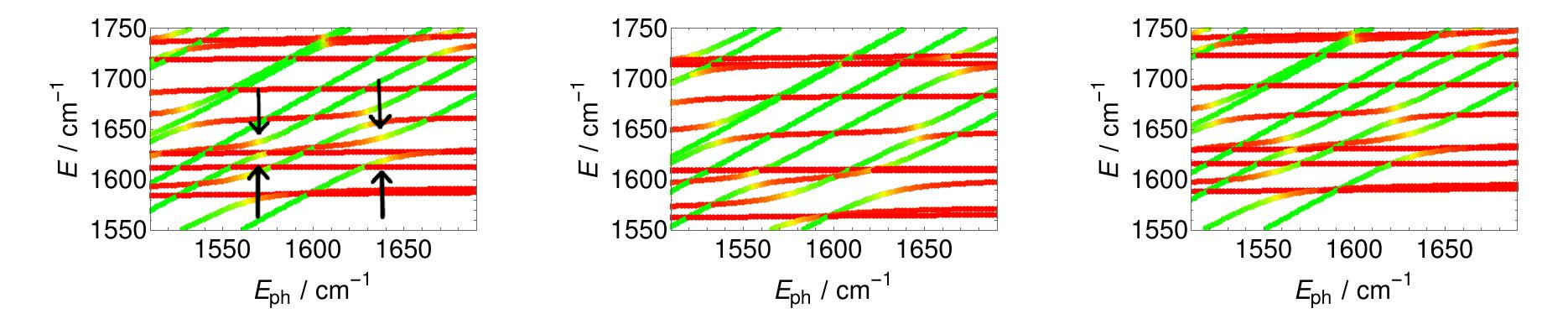}
\centering \includegraphics[width=1.00\textwidth]{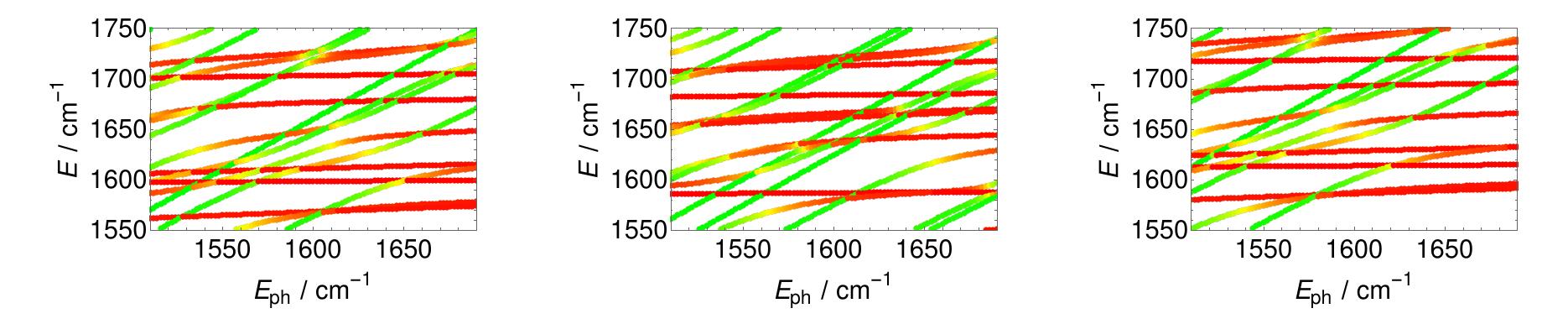}

\caption{Same as middle row of Figure \ref{fig:gvar_var_v2}, but the polaritonic energies are now expressed as a function of the $E_{\rm ph}$ cavity photon energy, at the coupling strengths of
$g=400$ \cm (upper row) and $g=800$ \cm (bottom row).
}
. \label{fig:nuvar_var_v2}
\end{figure}

Figure \ref{fig:nuvar_var_v2} presents the energy landscape for
$g=400$ \cm and $g=800$ \cm
coupling strengths and varying photon energy near the HOH bending fundamental.
In accordance with Figure \ref{fig:gvar_var_v2}, for
$g=400$ \cm  neglecting the polarizability interaction is an accurate approximation, however, for the larger coupling strength of
$g=800$ cm$^{-1}$,
this interaction also becomes important.
Figure \ref{fig:nuvar_var_v2} also demonstrates how different polaritonic states can be formed by a single  vibrational fundamental, depending on which rotational excitation the cavity photon is tuned to.
For example, in the upper left panel of Figure \ref{fig:nuvar_var_v2} two black arrows point out the polariton formation involving the $(0 1 0)[1 1 1]\leftarrow(0 0 0)[2 0 2]$ transition near $E_{\rm ph}=1570$ \cm and the polariton formation involving the $(0 1 0)[1 1 1]\leftarrow(0 0 0)[0 0 0]$ transition near $E_{\rm ph}=1630$ cm$^{-1}$.

\clearpage

\begin{figure}[h!]
\centering \includegraphics[width=0.80\textwidth]{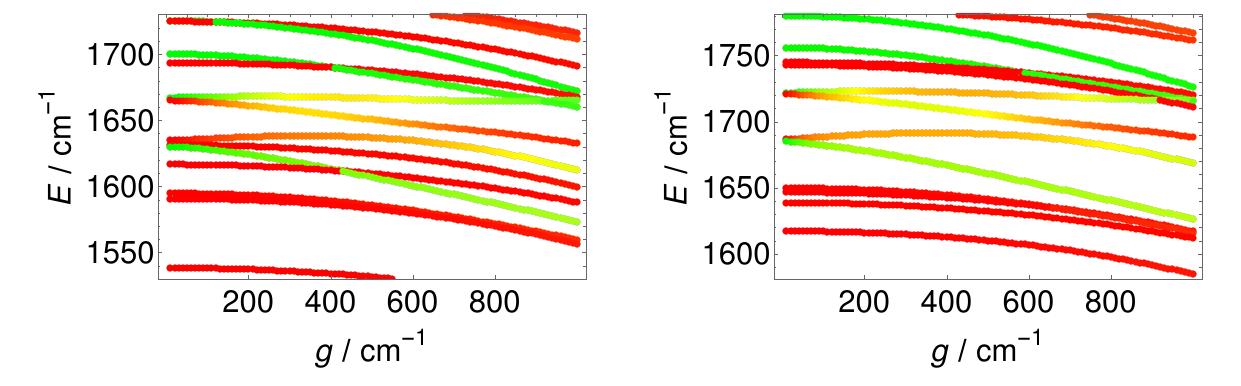}
\centering \includegraphics[width=0.80\textwidth]{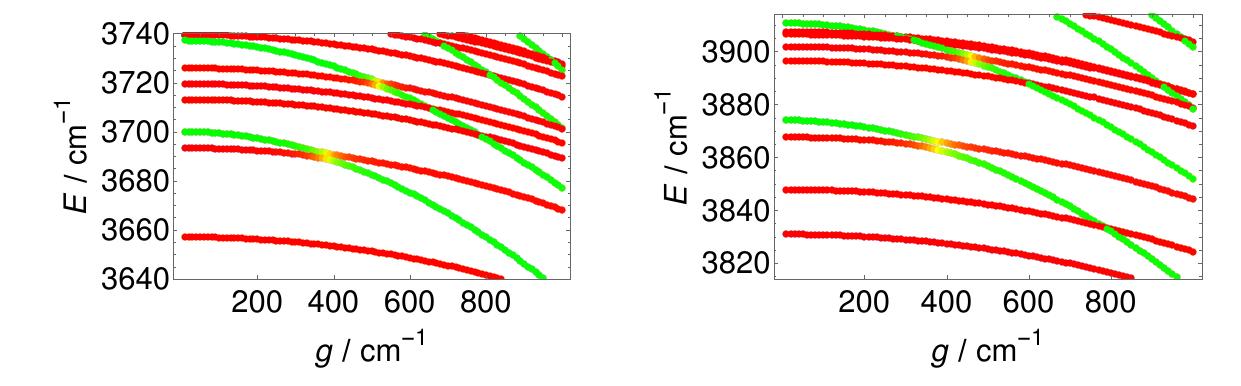}
\centering \includegraphics[width=0.80\textwidth]{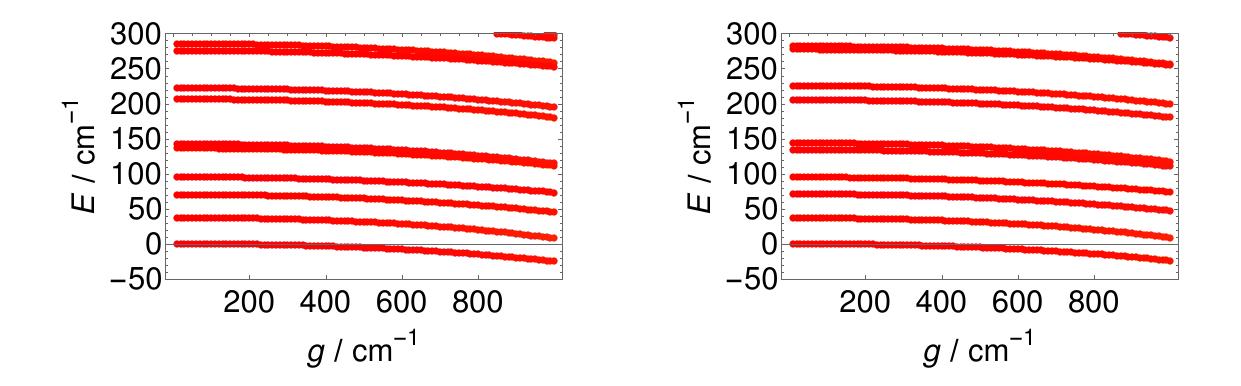}

\caption{Left column: converged eigenvalues of the rovibrophotonic Hamiltonian of Eq. (\ref{eq:hamiltonian_in_general_basis}) as a function of the $g$ coupling strength, obtained using the numerically exact variational molecular eigenstate basis and photon wave numbers of $\tilde{\nu}=1630$ cm$^{-1}$ (upper and lower panels) and $\tilde{\nu}=3700$ cm$^{-1}$ (middle panel).
Right column: same as in left column, but obtained using the HORR approximation for the molecular eigenstates and photon wave numbers of $\tilde{\nu}=1681$ cm$^{-1}$ (upper and lower panels) and $\tilde{\nu}=3874$ cm$^{-1}$ (middle panel).
}
. \label{fig:gvar_var_vs_HORR}
\end{figure}

Now we turn to the issue of the model used to represent the molecular rovibrational motion.
Figure (\ref{fig:gvar_var_vs_HORR}) compares the rovibropolaritonic energies obtained using the variational or HORR molecular models.
As seen in the upper and middle rows, for the semi-rigid H$_2$O molecule, the HO approximation works remarkably well for describing the coupling strength-dependence of the polaritonic eigenstates correlating with the vibrational fundamentals.
There is a shift in energy between the variational and HO vibrational fundamentals (51 cm$^{-1}$ in the upper row and 174 cm$^{-1}$ in the middle row), which can be easily taken care of by scaling the HO force constants \cite{Pulay1983}, a common procedure in theoretical molecular spectroscopy.
In addition, several red lines can be seen, which are located differently with respect to the vibrational fundamentals in the two columns.
These lines correspond to rotationally highly excited states, indicating that the tens of cm$^{-1}$ differences observed are a consequence of the RR approximation breaking down for rotationally highly excited states, as expected.
For low rotational excitations the two columns show nearly identical results, as presented by the lower row of Figure \ref{fig:gvar_var_vs_HORR}.
Overall, Figure \ref{fig:gvar_var_vs_HORR} implies that if the HORR approach is suitable for a specific system under field-free conditions, then it should be suitable to describe the light contaminated states in the IR microcavity as well.
Caution, however, is needed to make sure that the HORR approach is appropriate for the full energy range and degree of excitations
considered.

\subsection{Thermochemistry of H$_2$O in IR microcavities}

\begin{figure}[H]
\centering \includegraphics[width=0.4\textwidth]{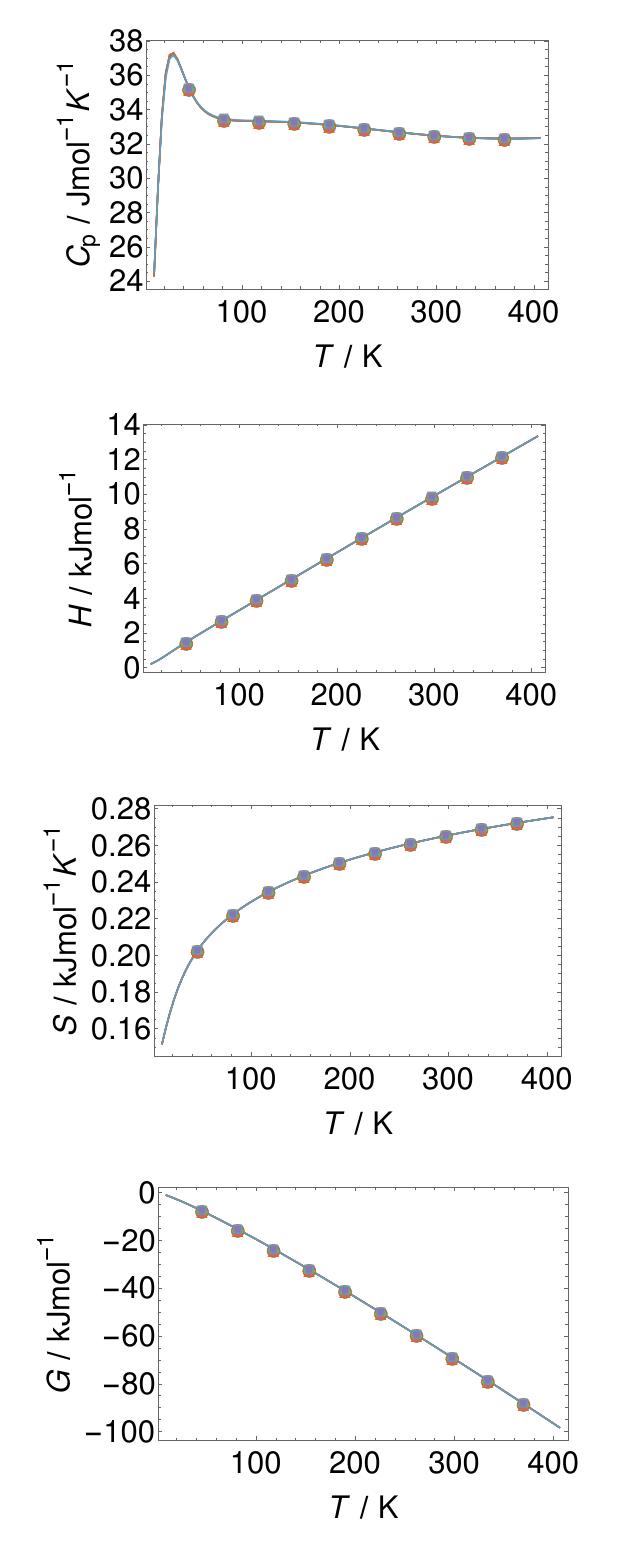}\centering
\centering \includegraphics[width=0.4\textwidth]{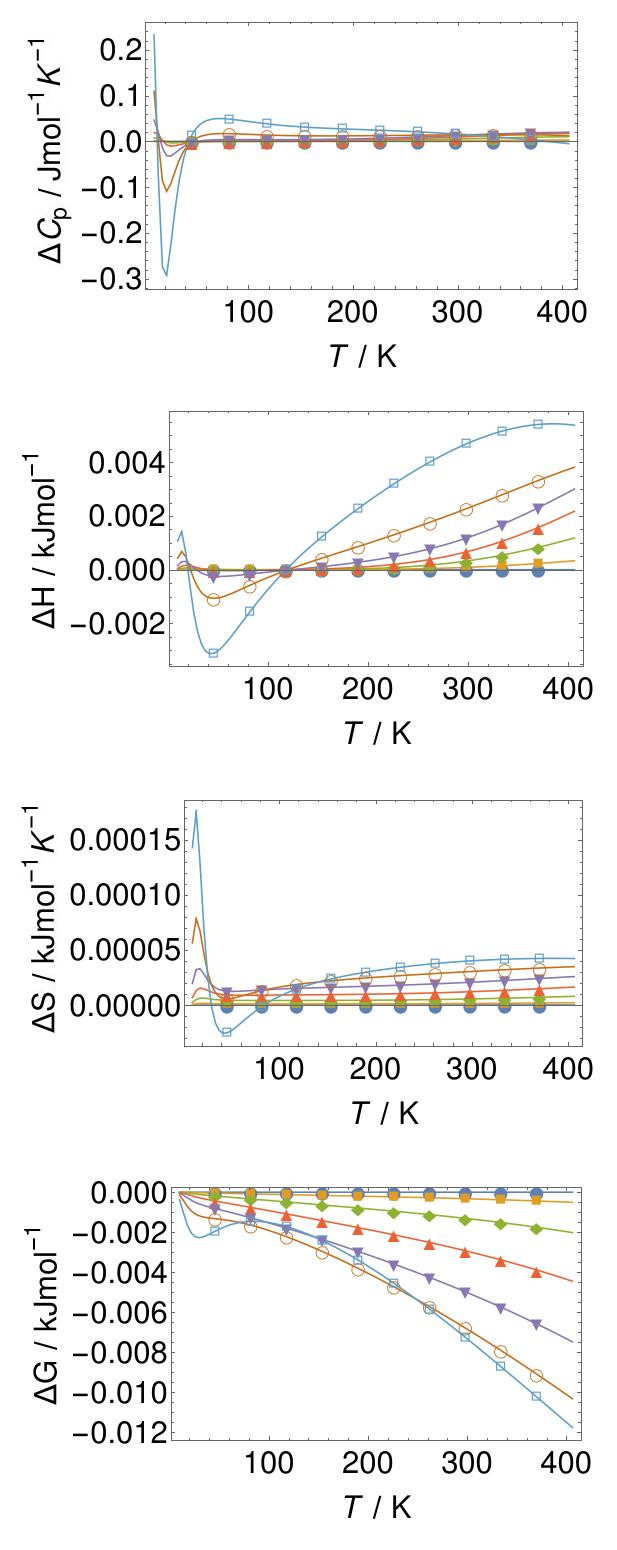}\centering
\includegraphics[width=0.15\textwidth]{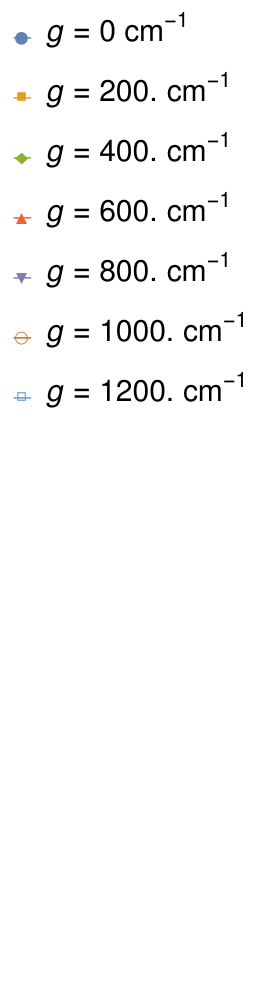}

\caption{
Various thermodynamic quantities presented as a function of temperature using $\tilde{\nu}=1630$ cm$^{-1}$ and different $g$ coupling strengths (left column), and the changes in the various thermodynamic quantities with respect to the $g=0$ scenario.
}
. \label{fig:gvar_var_thermo}
\end{figure}

With the rovibropolaritonic energy levels computed and available, one can construct the partition function
and derive other thermochemical quantities of the ``rovibrating H$_2$O + IR cavity mode" system at different temperatures \cite{16FuSzHrKy,Pilar2020}.
Note that using direct summation with the energy levels computed in this work yields results that do not include any effects of collective coupling, therefore, the computed values should not be regarded as quantitative, but rather as values that can be used to test how various approximation affect thermochemistry.
The issue of how the results presented here are related to the realistic mesoscopic systems with many water molecules collectively interacting with the radiation mode is left for future work.
In the following, results computed for ortho water are presented.
Figure \ref{fig:gvar_var_thermo} shows various thermodynamic quantities presented as a function of temperature at different $g$ coupling strengths for $\tilde{\nu}=1630$ cm$^{-1}$ photon wave number.
To highlight the changes induced by the light-matter coupling, Figure \ref{fig:gvar_var_thermo} also shows the changes in the various thermodynamic quantities with respect to the $g=0$ scenario.
It can be seen in Figure \ref{fig:gvar_var_thermo} that light-matter interaction changes the thermodynamic properties only to a very small extent, and the temperature dependence of these changes is not trivial.
In most cases the effect is negligible, the largest change of about one percent is seen for the heat capacity at low temperatures.

\begin{figure}[H]
\includegraphics[width=0.4\textwidth]{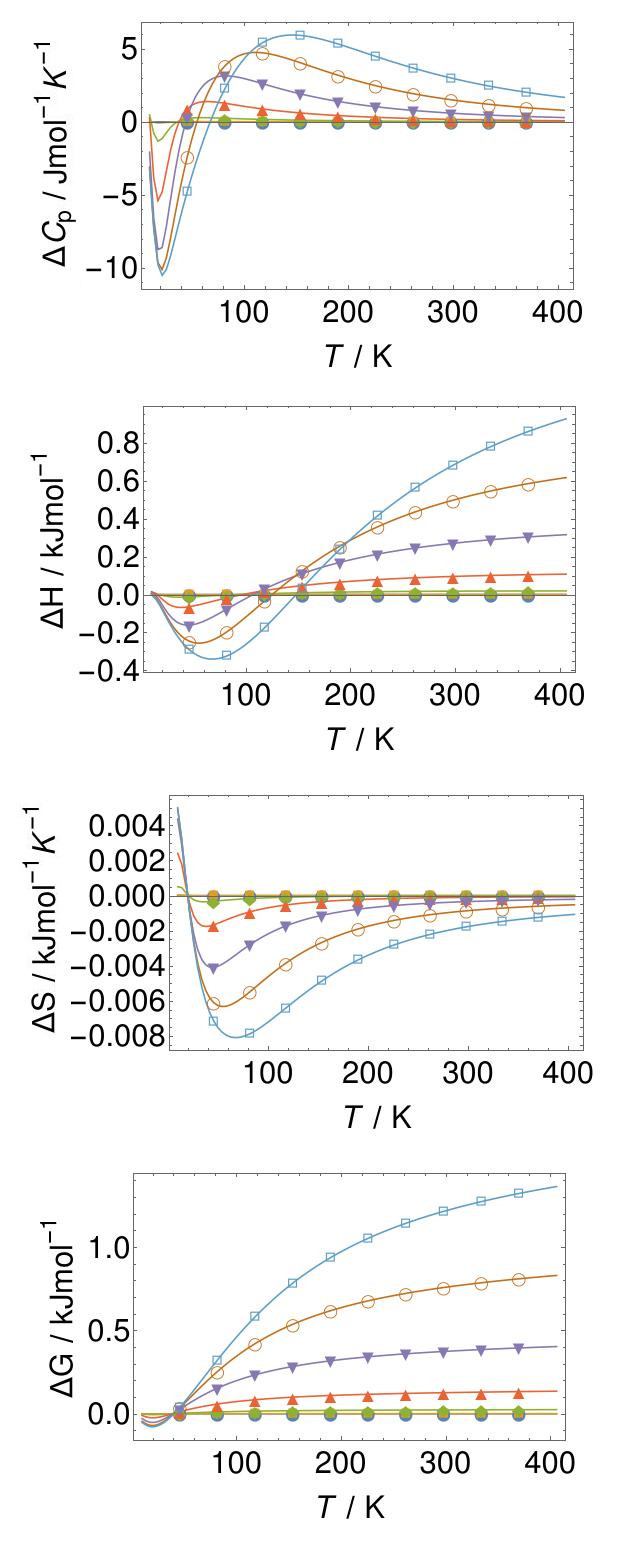}\centering
\centering \includegraphics[width=0.4\textwidth]{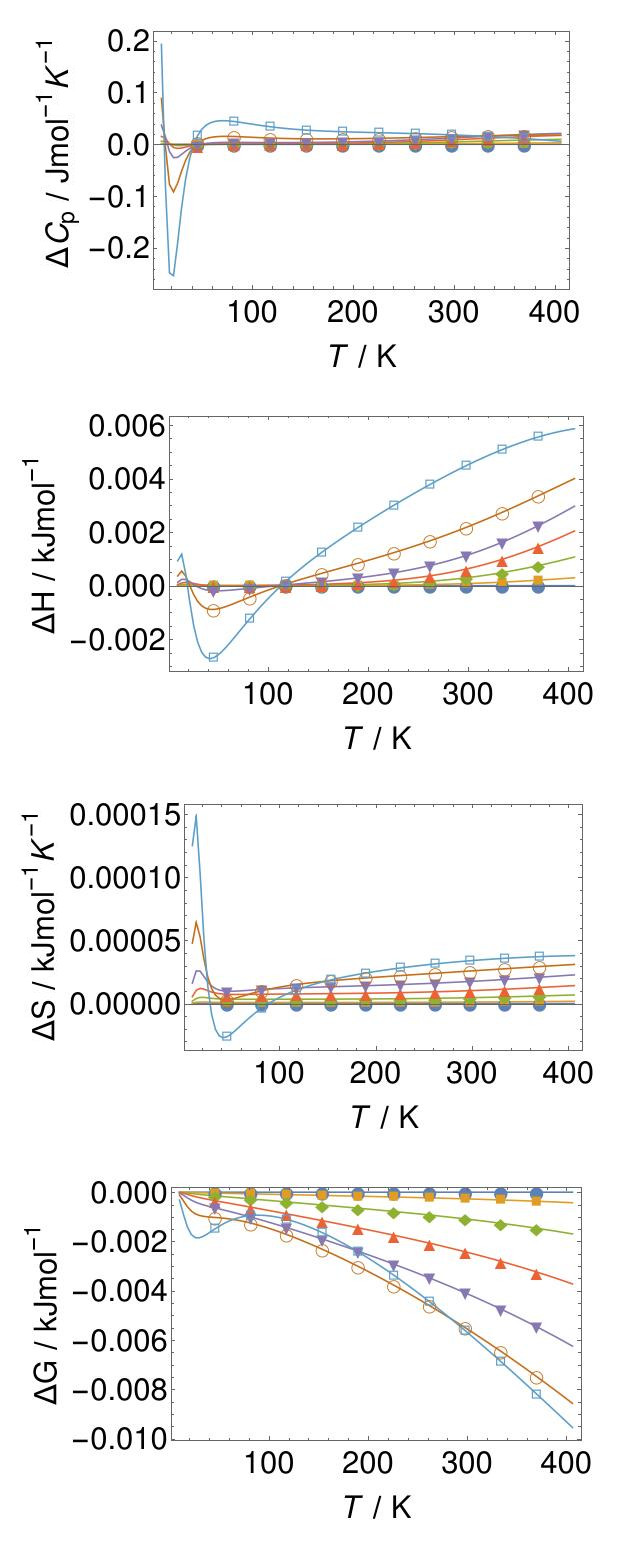}\centering
\includegraphics[width=0.15\textwidth]{gvar_gridplot_legend}

\caption{
Changes in the various thermodynamic quantities with respect to the $g=0$ scenario. Left column: $\tilde{\nu}=1630$ cm$^{-1}$, using the variational molecular basis, self dipole and polarizability interaction terms neglected. Right column: $\tilde{\nu}=1681$ cm$^{-1}$, using the HORR molecular basis and the full Hamiltonian of Eq. (\ref{eq:hamiltonian_in_general_basis}).
}
\label{fig:gvar_onlyDip_HORR_thermo}
\end{figure}

\begin{figure}[H]
\centering \includegraphics[width=0.8\textwidth]{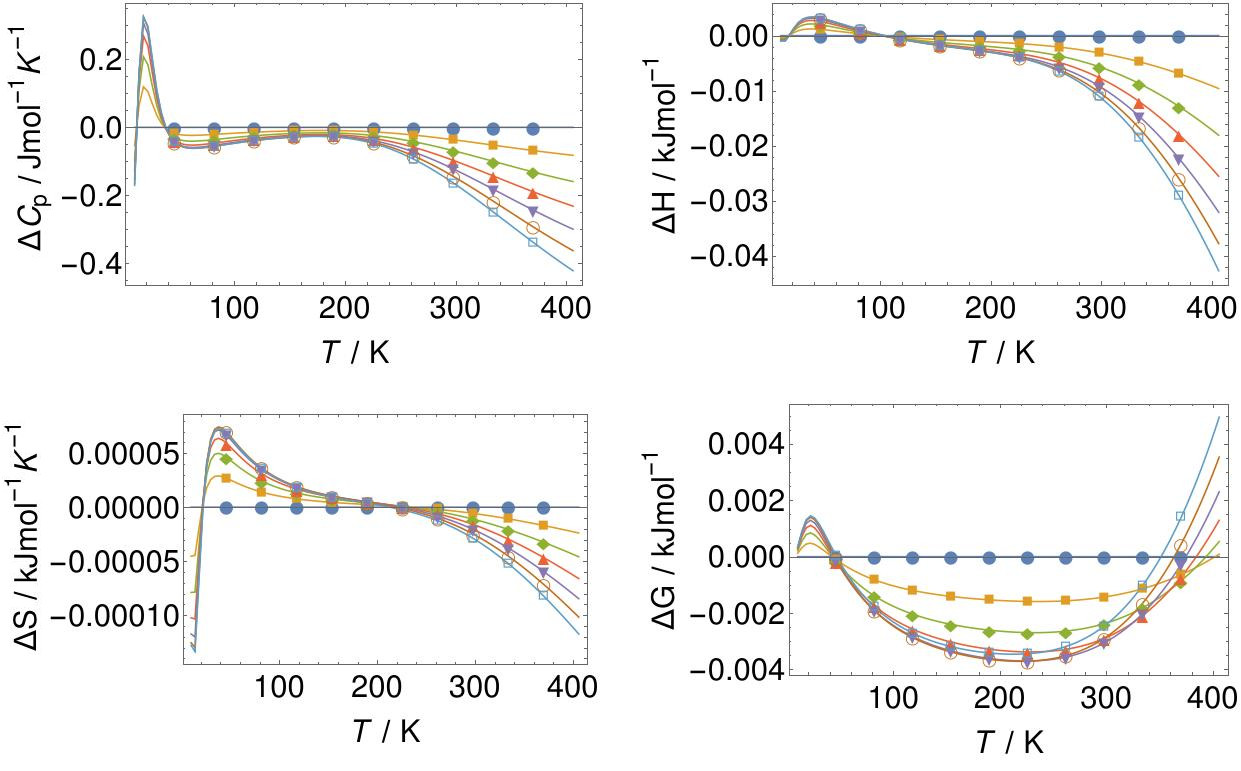}
\centering \includegraphics[width=0.18\textwidth]{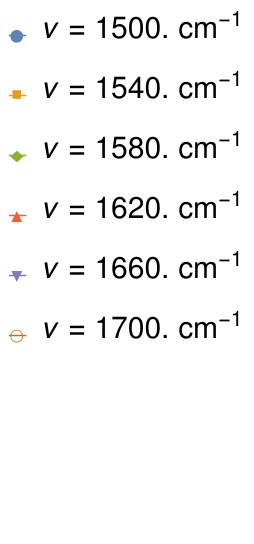}

\caption{
Changes in the thermodynamic quantities with respect to the $\tilde{\nu}=1500$ cm$^{-1}$ scenario using the full Hamiltonian of Eq. (\ref{eq:hamiltonian_in_general_basis}) with a fixed
$g=800$ \cm
value, presented as a function of temperature for different $\tilde{\nu}$ photon wave numbers near the HOH bending fundamental (exact resonance for $(0 1 0)[1 1 1]\leftarrow(0 0 0)[0 0 0]$ is at 1635.0 cm$^{-1}$).
}
. \label{fig:nuvar_var_thermo}
\end{figure}

As expected from Figure \ref{fig:gvar_var_v2} and shown in Figure \ref{fig:gvar_onlyDip_HORR_thermo}, including molecular polarizability and the self-dipole interaction is also necessary for thermochemistry, the permanent dipole-only model significantly overestimates the light-matter coupling-induced changes (compare right column of Fig. \ref{fig:gvar_var_thermo} with left column of Fig. \ref{fig:gvar_onlyDip_HORR_thermo}).
This is probably due to the large negative shift in the energy levels when the self-dipole term is negleted, see middle panels of Figures \ref{fig:gvar_var_v2}.
The right column of Figure \ref{fig:gvar_onlyDip_HORR_thermo} shows that for the summation of energy levels at ambient temperatures, the HORR approach performs with good accuracy, the cavity-induced changes in thermochemistry are nearly identical to those obtained with the variational molecular model.
Finally, Figure \ref{fig:nuvar_var_thermo} shows how various thermodynamic quantities change with respect to the $g=0$ scenario, using a fixed
$g=800$ \cm
value, presented as a function of temperature and for different $\tilde{\nu}$ photon wave numbers near the HOH bending fundamental.
Figure \ref{fig:nuvar_var_thermo} demonstrates systematic changes with respect to the photon wave number, however, no resonance effect is apparent.
This indicates that the efficient formation of rovibrational polaritons for specific vibrational modes does not dominate the thermochemistry derived from states localized near the PES minimum.
It is noted that while the absence of resonance effects in related works on reaction kinetics \cite{Galego_2019,CGA_2022} is a consequence of the model limitations, it is not necessarily an artifact of the theoretical approach in this work.


\section{Summary}
(1) A new theoretical framework was presented for the computation of molecular rovibrational polaritons in a lossless infrared (IR) cavity, in which the quantum description of molecular rotation and vibration can be formulated using arbitrary approximations.
The molecular parameters required by the proposed approach can be obtained from standard quantum chemistry tools, circumventing the need to rely on theoretical methods which explicitly incorporate the cavity radiation for determining molecular electronic properties.
The validity of the presented framework is justified as long as the
molecule can be described within the Born-Oppenheimer (BO) approximation, and the light-induced changes in the electronic structure can be treated perturbatively through molecular polarizabilities, i.e., the radiation field is not resonant with electronic transitions and the ultrastrong coupling regime is not reached.
(2) The rovibrational polaritons of H$_2$O were computed using different molecular models and considering the impact of different terms in the Hamiltonian of Eq. (\ref{eq:hamiltonian_in_general_basis}).
(2.1) Due to the light-induced energy shift of the molecular energy levels, resonance condition and the formation of polaritons is determined by the coupling strength and the photon energy simultaneously.
The energy shifts are dominantly negative, and depending on the relative polarizability of the molecule in the photon-excited state with respect to the rovibrationally excited state,
the efficient generation of polaritons at low coupling strengths could require the photon energy to be either red- or blue-shifted with respect to the field-free molecular transition.
(2.2) With increasing coupling strength $g$, the effects of the self-dipole interaction and the molecular polarizability increase, and need to be accounted for to obtain accurate rovibropolaritonic energy levels.
(2.3) For H$_2$O and in the investigated energy range the HORR approximation performs well in describing rovibropolaritonic properties, indicating that as long as the rovibrational model is appropriate for describing the field-free molecule, the computed rovibropolaritonic properties can be expected to be accurate as well.
(3) Thermochemical quantities were computed from the rovibropolaritonic energy levels of H$_2$O, neglecting collective effects.
It was found that strong light-matter coupling between the radiation mode of an IR cavity and the rovibrational states of H$_2$O lead to
very small
changes in the thermodynamic properties of the system, and these changes seems to be dominated by non-resonant interactions between light and matter.
Including molecular polarizability and the self-dipole interaction is very important, as the permanent dipole-only model significantly overestimates the light-matter coupling-induced changes in thermochemistry.

\section{Acknowledgements}
I thank anonymous Referee 2 for the valuable comments that considerably improved the publication.
This research was supported by the János Bolyai Research Scholarship of the Hungarian Academy of Sciences and by the ÚNKP-22-5 New National Excellence Program of the Ministry for Innovation and Technology from the source of the National Research, Development and Innovation Fund.
The Author is also grateful to NKFIH for additional support (Grant No. FK134291).

\clearpage

\bibliography{main}

\end{document}